\documentclass[dvips]{article}
\usepackage{icrctc07}

\def\arcmin{\hbox{$^\prime$}}
\def\arcsec{\hbox{$^{\prime\prime}$}}

\title{New Companions for the lonely Crab?\\
VHE emission from young pulsar wind nebul{\ae} revealed by H.E.S.S.
}
\shorttitle{New Companions for the lonely Crab?}
\authors{  A.~Djannati-Ata\"i$^{1}$, O.C.~de~Jager$^{2}, $
R.~Terrier$^{1}$,  Y.A.~Gallant$^{3}$ \& S.~Hoppe$^{4}$   
for the H.E.S.S. collaboration$^{5}$}
\shortauthors{A.~Djannati-Ata\"i et al}
\afiliations{$^1$ APC (CNRS, Universit\'e Paris VII, CEA, Observatoire de Paris), Paris, France \\ $^2$ Unit for Space Physics, North-West University, Potchefstroom 2520, South Africa\\ $^3$ LPTA,  Universit\'e Montpellier 2, IN2P3/CNRS,, Montpellier, France \\ $^4$Max-Planck-Instituit f\"ur Kernphysik, PO box 103980, D69029 Heidelberg, Germany\\ $^5$ \rm{ \texttt{www.mpi-hd.mpg.de/HESS}}
}
\email{djannati@apc.univ-paris7.fr}

\abstract{The deeper and more extended survey of the central parts of the 
Galactic Plane by H.E.S.S. during 2005-2007 has revealed a number 
of new point-like, as well as, extended sources. Two point-like sources
can be associated to two remarkable objects around ``Crab-like'' young
and energetic pulsars in our Galaxy : G21.5-0.9 and Kes~75. The
characteristics of each of the sources are presented and possible
interpretations are briefly discussed.}

\begin{document}
\maketitle

\section{Introduction}

\begin{figure*}[!t]
\begin{center}
\includegraphics*[width=0.46\textwidth,angle=0,clip]{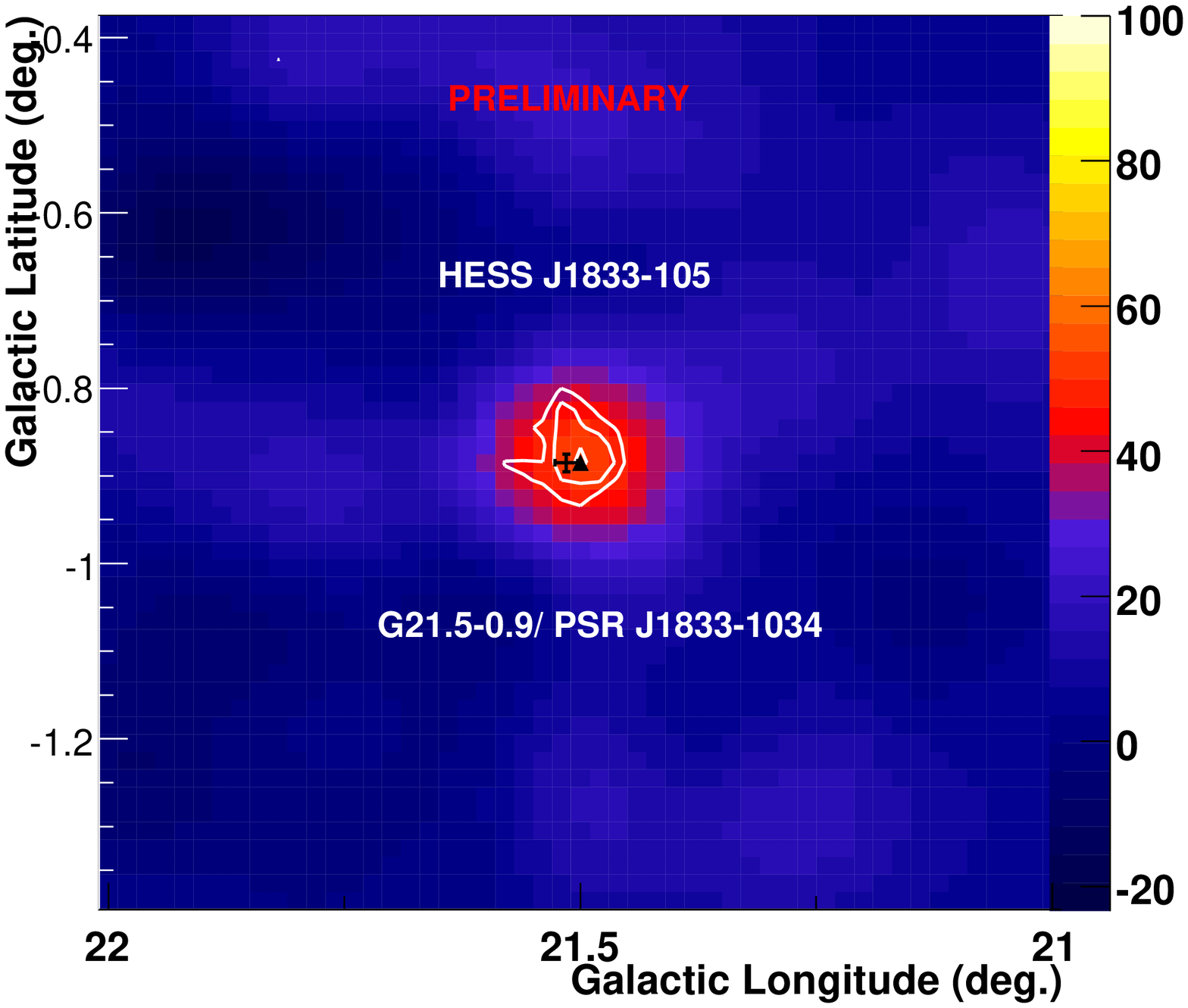}
\includegraphics*[width=0.46\textwidth,angle=0,clip]{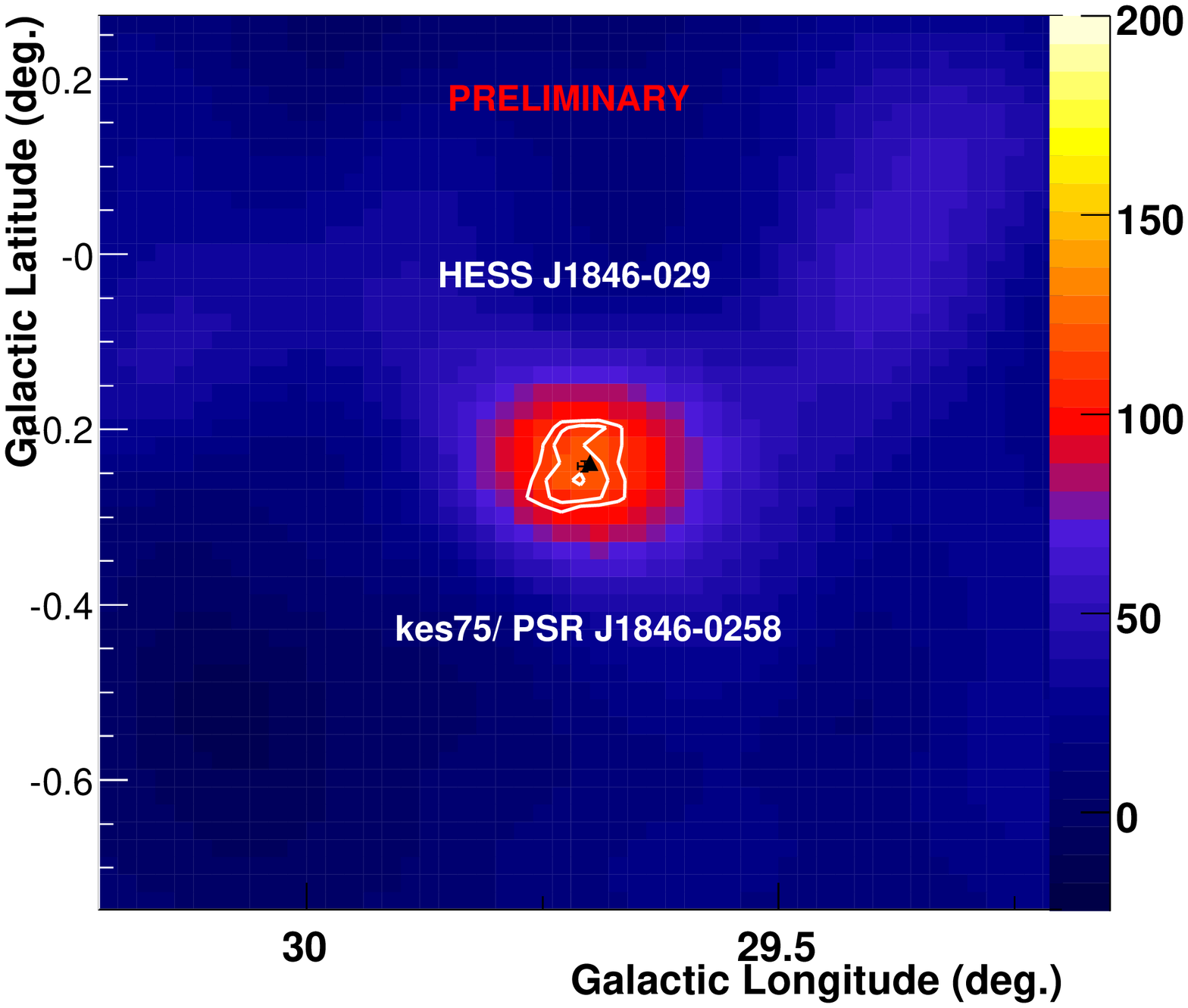}
\end{center}
\caption{ Smoothed excess maps
($\sigma=0.08^{\circ}$) of the 0.5$^{\circ}\times0.5^{\circ}$ field of
view around the positions of HESS~J1833-105 (left) and HESS~J1846-029
(right). The white contours show the pre-trials significance levels for 4, 5,
6 $\sigma$, and 7, 8, 9 $\sigma$, respectively. The black triangle
marks the position of the pulsars. The best-fit positions of the
  two sources are marked with an error cross (for HESS~J1846-029 the
latter overlaps with the triangle).}
\label{skymaps}
\end{figure*}

The standard candle of VHE astronomy, the
Crab Nebula, has served for decades as a yardstick in almost all
wavelengths, and yet it is a very peculiar object, harbouring the most
energetic and one of the youngest pulsars of our Galaxy.
Since the early days, where the similarities of the historical trio Crab/3C~58/G~21.5-0.9
were under debate~\cite{WilsonWeiler76}, radio and X-ray astronomy have
provided a wealth of information by detecting and characterizing
nebulae around rotation-powered pulsars.
In the VHE domain, H.E.S.S. has revealed
more than a dozen pulsar wind nebulae (PWN), either firmly established
as such or compelling candidates~\cite{HESSpwnICRC07}, almost all of
which are middle-aged (at least few kyrs up to $\sim$100 kyrs, except
MSH~15-52) and exhibit an offset between the pulsar position and the
nebula center.
We report here on the VHE emission discovery of two remarkable
objects, G~21.5-0.9 and Kes75, which also harbor very young and
energetic pulsars and which on some aspects, especially their
plerionic nebular emission due to an energetic pulsar, can be
considered as Crab-like.

\begin{figure*}[tb] 
\begin{center}
\includegraphics[width=0.48\textwidth]{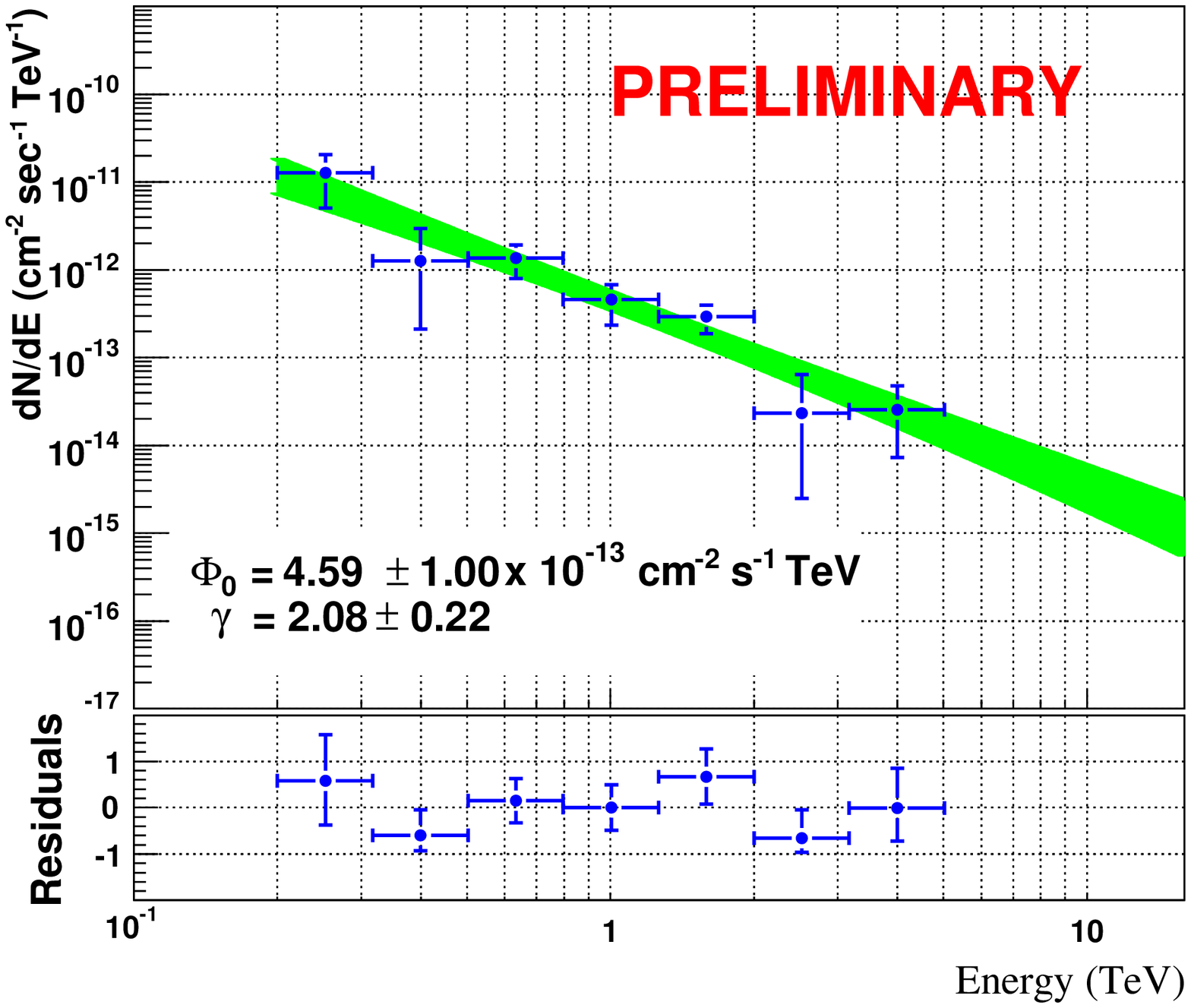}
\includegraphics[width=0.48\textwidth]{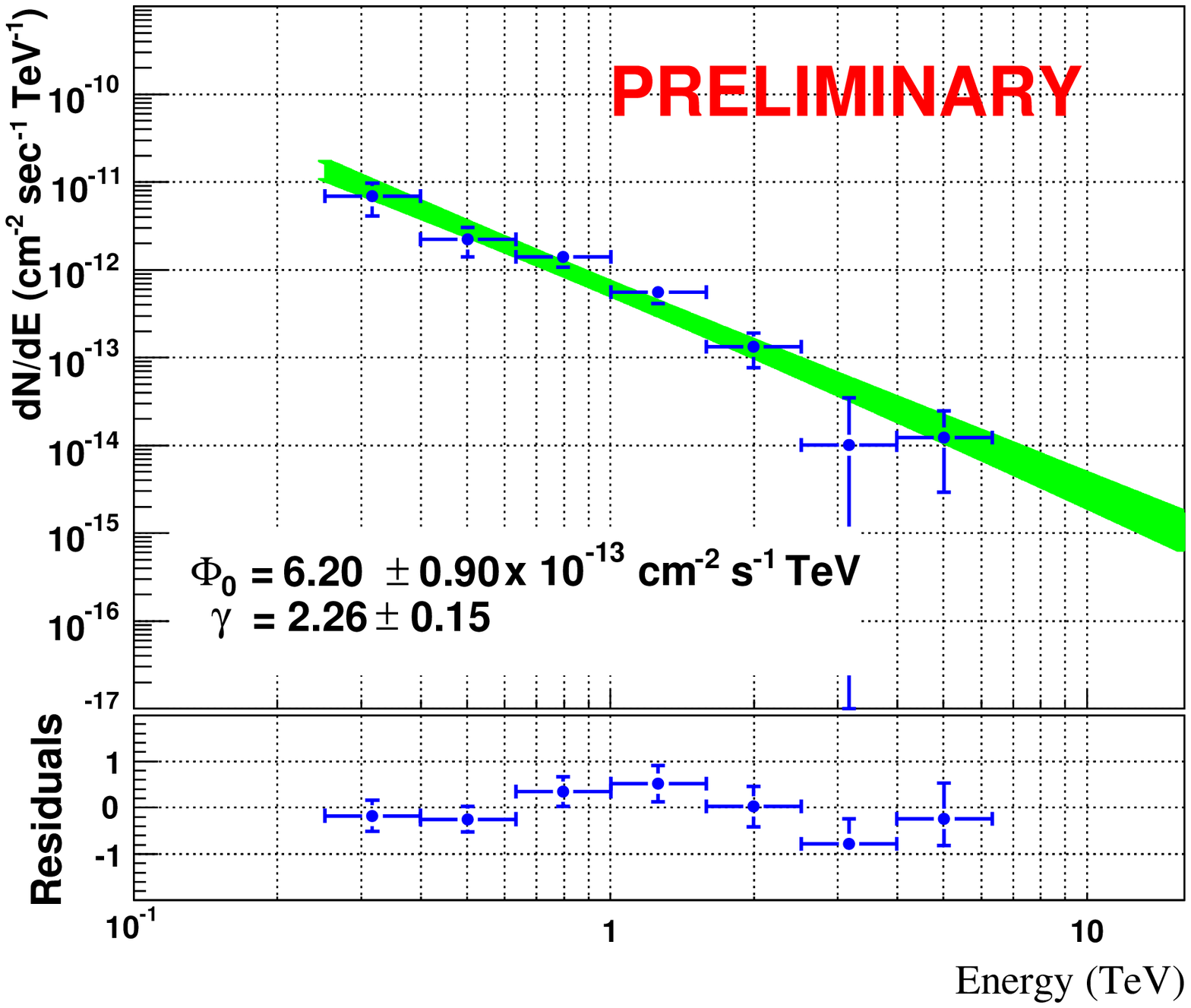}
\end{center}
\caption{Differential energy spectra above for HESS~J1833-105 (left) and HESS~J1846-029
(right). The shaded area shows the 1 $\sigma$
confidence region for the fit parameters.} 
\label{spectra} 
\end{figure*} 

G21.5-0.9 \cite{Altenhoff70}, recently revealed as a composite SNR
consisting of a centrally peaked PWN and a 4{\arcmin}
shell~\cite{Bocchino2005,MathesonSafi-Harb2005}, was 
previously classified as one
of the about ten Crab-like SNR~\cite{Green2004}. Its flat spectrum
PWN is polarised in radio~\cite{BeckerSzymkowiak1981} with a spectral
break above 500 GHz~\cite{GallantTuffs1998}. The non-thermal X-ray PWN 
with radius $\sim$40{\arcsec} shows significant evidence of cooling
~\cite{Slane2000}, with the power-law photon index steepening from
1.43$\pm$0.02 near the pulsar to 2.13$\pm$0.06 at the edge of the 
PWN. 
There appears to be a synchrotron X-ray halo at a radius of 140{\arcsec} from the
pulsar which could originate in the shell~\cite{Bocchino2005,MathesonSafi-Harb2005},
with a contribution of scattering off dust grains
as proposed by Bocchino et
al.~\cite{Bocchino2005}. 
The 61.8 ms pulsar PSR~J1833-1034, with a spin-down power
of ${\dot E} = 3.3\times 10^{37}$erg/s and a characteristic age of
4.9 kyr was discovered only recently through its faint radio pulsed
emission~\cite{Gupta2005,Camilo2006}. 
Given the derived distance of 4.7$\pm$0.4 kpc, the age of G~21.5-0.9
was revised downwards by a factor of $\sim$10 
to force consistency with the freely expanding
SNR shell~\cite{Camilo2006}.  PSR~J1833-1034 in G~21.5-0.9 is the second most energetic
pulsar known in the Galaxy.

Kes 75 (SNR G29.70.3) is also a prototypical example of a composite
remnant for which the distance of 19 kpc was estimated
through neutral hydrogen absorption measurements~\cite{BeckerHelfand1984}.
Its 3.5{\arcmin} radio shell surrounds a flat-spectrum highly
polarized radio core, 
and harbors, at its
center, the 325 ms X-ray Pulsar, PSR~J1846-258~\cite{Gotthelf2000}. 
The latter has the shortest known characteristic age $\tau_c= 723$ yr
and a large inferred magnetar-like magnetic field of B$=4.9\times
10^{13}$G. The pulsar lies within a 25{\arcsec}$\times$20{\arcsec} X-ray
nebula which exhibits an 
photon index of 1.92$\pm$0.04, but no evidence of cooling as a function
of the distance to the pulsar.
Like in G~21.5-0.9 there is an X-ray halo, in this case due mostly to dust
scattering, but a non-thermal contribution 
from electrons accelerated in the shell remains possible~\cite{Helfand2003}.

\section{Observations, Analysis \& Results}
\label{results}

Results presented in this section should be
considered as preliminary.  

The first H.E.S.S. observations of G~21.5-0.9 and Kes~75 were
performed during 2004 and 2005 as part of the systematic survey of the
inner Galactic plane within the longitude range
$ l\in$[$-30^{\circ}$,$+30^{\circ}$] and latitude band 
$b\in$[$-3^{\circ}$,$+3^{\circ}$]. Kes~75, at the edge of the first survey,
was covered in the extension to $l\in$[$+30^{\circ}$,$+60^{\circ}$] of
the survey in the years 2005-2007. The data obtained through the systematic 
survey was completed with followup observations of promising
candidates in wobble mode, hence the two sources are offset at various angular distances with
respect to the center of the field of view. 
The total quality-selected and dead-time corrected data-set includes
19.7 hours of data on G~21.5-0.9 and 24.1 hours on Kes~75, with average offsets of
1.33$^{\circ}$ and 1.1$^{\circ}$, for each source, respectively.

The standard scheme for the reconstruction of events was applied to
the data \cite{HESSCrab}. Cuts on the scaled width and length of images (optimised 
on $\gamma$-ray simulations and off-source data) were used to suppress the
hadronic background. As described e.g. in
\cite{HESSKooka}, sky-maps and morphological analyses are made with a
tight cut on the image size of 200 p.e. (photoelectrons) to achieve a
maximum signal-to-noise ratio and a narrow PSF (point spread
function). For the spectral analysis,
the image size cut is loosened to 80 p.e. in order to cover the
maximum energy range. The background estimation for each position in the
two-dimensional sky map is made in the same way as for search of
extended sources \cite{HESSJ1908ICRC07}, i.e. computed from a ring
with a radius of $1.0^{\circ}$. For a point-like source this radius yields  
seven times a larger area for the background estimation
than the on-source region. 
The background used for the derivation of the spectrum, is evaluated
from circular regions in the field of view with the 
same radius and same offset from the pointing direction as that of the source region.
Finally, to avoid contamination of the background, events coming from
known sources were excluded.

Fig.~\ref{skymaps} shows the Gaussian-smoothed excess maps for
HESS~J1833-105 and HESS~J1846-029 where the white contours mark the
pre-trials significance levels. Both sources were first discovered as 
hot-spots within the analysis scheme described above and then
confirmed through additional followup data at pre-trials significance
of 6.4 and 9.9 standard deviations, respectively. 
A conservative estimate of the trials yields post-trials
significance of 4.0 $\sigma$ and 8.3 $\sigma$ for HESS~J1833-105 and
HESS~J1846-029, respectively.

 
The extension and the position of the sources were evaluated by adjusting to the images
a symmetrical two-dimensional Gaussian
function, convolved with the instrument PSF (5{\arcmin} for this
analysis). The best-fit positions lie at 
18$^{\rm h}$33$^{\rm m}$32.5$^{\rm s}$$\pm$0.9$^{\rm s}$,$-$10d33\arcmin 19\arcsec$\pm$55\arcsec and 
18$^{\rm h}$46$^{\rm m}$24.1$^{\rm s}$$\pm$0.5$^{\rm s}$,$-$02d58\arcmin 53\arcsec$\pm$34\arcsec. 
The intrinsic extensions are compatible with a point-like source for both
sources and their positions are in a quite good agreement with the pulsars
associated to each supernova remnant, i.e. PSR~J1833-1036
(18$^{\rm h}$33$^{\rm m}$33.57$^{\rm s}$,$-$10d34\arcmin 7.5\arcsec) 
and PSR~J1846-0258 (18$^{\rm h}$46$^{\rm m}$24.5$^{\rm s}$,$-$02d58\arcmin 28\arcsec).

The energy spectra of the two sources are derived using the
forward-folding maximum likelihood 
fit of a power-law~\cite{CATSpectrum}. The fluxes are at a level of
$\sim$2\% of that of the Crab Nebula and the spectra are rather hard
(Fig.~\ref{spectra}):  the photon indices are $2.08 \pm
0.22_{\rm stat}$ and $2.26 \pm0.15_{\rm stat}$ for  HESS~J1833-105 and
HESS~J1846-029, respectively, with a systematic error of $\pm 0.1$.

\section{Discussion}

It is remarkable that de Jager et al.~\cite{deJager1995} predicted 
that plerionic VHE $\gamma$-rays from G21.5-0.9 would be detectable
at a level of $4\times10^{-13}$ cm$^{-2}$~s$^{-1}$ at 1~TeV with an electron spectral index
of $\sim$2.8, which would give a photon index near 2.0 at VHE energies (after including KN
effects given the contributions from dust and CMBR). 
Their prediction was based on an assumed equipartition field strength
of 22 $\mu$G which is close to the value of $\sim$15$\mu$G implied
from $\gamma$-ray observations reported here (assuming IC scattering on CMB photons
only, and using the ratio of the X-ray to the $\gamma$-ray
luminosities: $L_{\rm X}/L_{\gamma}\sim30$). The equipartition field strength was afterwards
increased to 0.3 mG following the revision of the maximum spectral range of the radio PWN
to 500 GHz \cite{GallantTuffs1998, Camilo2006}. However, the detection of VHE
$\gamma$-rays by H.E.S.S. from PWN tends to confirm the suggestion of
Chevalier \cite{Chevalier2004} that some PWN may be particle
dominated, so that the true PWN field strength may be significantly
lower than equipartition for some objects. 
In the case of Kes~75, $L_{\rm X}/L_{\gamma}\sim$10
yields also a lower than equipartition nebular magnetic field
strength of $\sim$10 $\mu$G. It should be noted that Kes~75 shows the
highest conversion efficiency in X-rays ($\sim15$\%) as compared to other
``Crab-like'' pulsars ($\sim3$\% and $\sim0.6$\% for the Crab and
G~21.5-0.9, respectively) and a 100 times larger $\gamma$-ray efficiency
($\sim$2\%) than the Crab and G~21.5-0.9 which are similar in that
respect ($\sim$0.02\%).
However, the latter object's $L_{\rm X}/L_{\gamma}\sim30$ is 4 times
smaller than that of the Crab Nebula $L_{\rm X}/L_{\gamma}\sim120$. 
These numbers together with the spin
parameters and high surface magnetic field in the case of
PSR~J1846-0258, show that these objects, although ``Crab-like'' in
some aspects, do possess peculiar properities.

Given the evidence for synchrotron emission in the SNR
shell, an alternative interpretation of the VHE
emissions of G~21.5-0.9 and Kes75 would 
be radiation from particles accelerated at the non-relativistic forward
shock of the freely expanding SNR. However
the required field strength in the shell to explain the
H.E.S.S. detection in terms of IC scattering should be 
much lower than 10 $\mu$G, value which may be unreasonably low for
typical expanding SNR shells. 
Deeper observations of both sources
could help to constrain the size of the VHE emission region and to 
ascertain whether it is compatible with this scenario.

\section*{Acknowledgments}
The support of the Namibia authorities and of the University of Namibia
in facilitating the construction and operation of H.E.S.S. is gratefully
acknowledged, as is the support by the German Ministry for Education and
Research (BMBF), the Max Planck Society, the French Ministry for Research,
the CNRS-IN2P3 and the Astroparticle Interdisciplinary Programme of the
CNRS, the U.K. Particle Physics and Astronomy Research Council (PPARC),
the IPNP of the Charles University, the Polish Ministry of Science and
Higher Education, the South African Department of
Science and Technology and National Research Foundation, and by the
University of Namibia. We appreciate the excellent work of the technical
support staff in Berlin, Durham, Hamburg, Heidelberg, Palaiseau, Paris,
Saclay, and in Namibia in the construction and operation of the
equipment.

\bibliography{icrcYoungPulsars}
\bibliographystyle{plain}

\end{document}